\documentclass[aps,prd,twocolumn,amsmath,amsfonts,amssymb,groupedaddress,nofootinbib,nobalancelastpage,floatfix,
secnumarabic,superscriptaddress,secnumarabic,preprintnumbers]{revtex4-2}
\pdfoutput=1
\usepackage{epsfig, graphicx, hyperref, slashed,xspace}
\usepackage[dvipsnames]{xcolor}
\usepackage[utf8]{inputenc}
\usepackage[T1]{fontenc}
\usepackage{cleveref}
\usepackage{soul}
\usepackage{diagbox}
\usepackage{booktabs}
\usepackage{multirow}
\usepackage{setspace}
\usepackage{float}
\usepackage{hyperref}
\usepackage{xurl}   % allows URL line breaking
\hypersetup{
  colorlinks=true,
  linkcolor=blue!60!black,
  citecolor=green!50!black,
  urlcolor=blue!70!black
}

\hypersetup{colorlinks=true,linkcolor=Maroon,citecolor=ForestGreen,filecolor=ForestGreen,urlcolor=ForestGreen}
\usepackage[most]{tcolorbox}
\usepackage{xspace}

\newcommand{\MG}{\textsc{MadGraph5\_aMC@NLO}\xspace}
\newcommand{\Pythia}{\textsc{Pythia8}\xspace}

\newcommand{\Root}{\textsc{ROOT}\xspace}
\newcommand{\lhe}{\textsc{LHE}\xspace}
\newcommand{\event}{\textsc{EventLoop}\xspace}
\newcommand{\lhereader}{\textsc{LHEReader}\xspace}
\newcommand{\FastJet}{\textsc{FastJet}\xspace}
\newcommand{\MadAnalysis}{\textsc{MadAnalysis5}\xspace}

\makeatletter\def\l@subsubsection#1#2{}\makeatother
\usepackage{orcidlink}

\usepackage{tcolorbox}
\tcbuselibrary{listings, breakable}

\newtcblisting{codebox}{
  colback=blue!5!white,
  colframe=blue!75!black,
  boxrule=0.8pt,
  arc=4pt,
  listing only,
  listing options={
    basicstyle=\ttfamily\small,    % slightly larger text
    breaklines=true,               % wrap long lines
    postbreak=\mbox{\textcolor{red}{$\hookrightarrow$}\space},
  },
  enhanced,
  breakable,
  left=2mm, right=2mm,     % horizontal padding
  top=2mm, bottom=2mm,     % vertical padding
  boxsep=2mm              % inner separation to prevent overflow
}

\newtcblisting{terminalbox}{
  colback=green!8!white,          % light green background
  colframe=green!50!black,        % darker green frame
  boxrule=0.8pt,
  arc=4pt,
  listing only,
  listing options={
    basicstyle=\ttfamily\small,
    breaklines=true,
    columns=fullflexible,
    keepspaces=true,
  },
  enhanced,
  breakable,
  left=2mm, right=2mm,
  top=2mm, bottom=2mm,
  boxsep=2mm
}

\usepackage{tikz}
\usepackage{tikz-feynman}

\tikzfeynmanset{compat=1.1.0}

\begin{document}

%\preprint{...}

\title{\mbox{\textsc{LHEReader}: Simplified Conversion from Les Houches Event Files to ROOT Format}}

\author{Aman Desai \orcidlink{0000-0003-2631-9696}}\email{aman.desai@adelaide.edu.au}
\affiliation{Department of Physics, Adelaide University, North Terrace, Adelaide, SA 5005, Australia}

\date{\today}

\begin{abstract}
We present the \textsc{LHEReader} program that converts a Les Houches Event file into a \Root file, allowing one to subsequently analyse the file using \Root. We evaluate the performance of this conversion by simulating $pp\to jj$ and $e^+e^- \to ZH$. The application of this program is illustrated through the simulation of $pp\rightarrow ZH$ with $Z\to\ell^+\ell^-$ and $H\to b\bar{b}$, analysing the events at the parton level after hard-scatter event generation as well as after the fast simulation stage using \MadAnalysis. The analysis is implemented in \Root, demonstrating the use of this program. The program is available at \url{https://github.com/amanmdesai/LHEReader}.

\end{abstract}

\maketitle

\section{Introduction}\label{sec:intro}

Monte Carlo event generators store parton-level event records using a standard file format called the Les Houches Event (\lhe) file~\cite{Alwall:2006yp}. These files store all necessary information about an event such as the particle identities, status codes, particle four-momenta, particle spin, information about the parent particles, and colour information. Among other variables, they store the scale, $\alpha_{\rm{QED}}$, $\alpha_{\rm{QCD}}$, and the event weight. The parton-level events generated, for example, by \MG~\cite{Alwall:2014hca} are usually passed to parton shower and hadronisation stages, which are handled by general-purpose event generators such as \Pythia~\cite{Bierlich:2022pfr}. Moreover, software such as \MadAnalysis~\cite{Conte:2012fm,Conte:2018vmg} can also produce its output in the form of an \lhe file. To facilitate the reading of these files, a variety of software tools have been proposed and are available for installation, see for example Ref.~\cite{lukas_heinrich_2025_17937243}.

In this paper, we present an alternative program, named \lhereader~\cite{aman_desai_2026_18822147}, to convert an \lhe file into a \Root file~\cite{Brun:1997pa}, which allows one to further analyse the parton-level events using functionalities defined within \textsc{CERN} \Root. We released version 1.0 of this program in February 2023~\cite{aman_desai_2023_7642763} and recently updated the program to allow additional input types such as \texttt{.lhe.gz}. In this paper, we also provide an \event code snippet which one can utilise to analyse files. Previously, we presented this program at the Australian Institute of Physics Congress 2024~\cite{AIPCongress2024}. The \lhereader has been used in research projects related to the LHC and FCC-ee cases~\cite{DeCristo2026HEPStuff,Nplastir2026LHEscripts,Cbell14022026MultiJetStudy,Hatakeyama2026POWHEG,ZBahariyoon2026FCCAnalyses,whizardLHEValidation2024}.

\begin{figure}[htbp]
    \centering
    \includegraphics[width=0.45\linewidth]{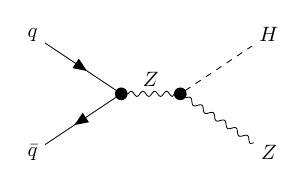}
    \caption{A representative leading-order Feynman diagram for $pp\to ZH$ production.}
    \label{fig:ZH_diagram}
\end{figure}

To illustrate the \lhereader program~\cite{aman_desai_2026_18822147} as well as \event, we simulate a Monte Carlo sample corresponding to $pp\to Z(\to \ell^+\ell^-) H(\to b\bar{b})$. The representative Feynman diagram for the production process is given in \autoref{fig:ZH_diagram}.

The organisation of this paper is as follows: in \autoref{sec:soft} we discuss the program, its dependencies and provide a quick start guide; in \autoref{sec:event} we provide the \event function and its usage description. Lastly, in \autoref{sec:mcsample} we apply the program on a physics example simulating $pp\to ZH$ process at the LHC and convert the final \lhe file into \Root and use the \event function on the physics sample for analysis.

\section{\lhereader program}\label{sec:soft}

The \lhereader program is available for use by directly installing the package from the following URL:

\begin{terminalbox}
https://github.com/amanmdesai/LHEReader
\end{terminalbox}

Alternatively, one can install this program through terminal by using the following command:

\begin{terminalbox}
git clone https://github.com/amanmdesai/LHEReader.git
\end{terminalbox}

The installation has been tested on a Linux-based system with the prerequisite that \textsc{CERN Root} software is available along with \textsc{PyRoot}~\cite{Galli:2020boj} enabled. The current version has been tested to work with \textsc{Root 6.36.04} and \textsc{Python 3.10}~\cite{python}.

The program is therefore available to perform conversion from an \texttt{.lhe} or \texttt{.lhe.gz} file to a \texttt{.root} file.

Assuming that an \lhe file, for example, \texttt{input.lhe}, is available in the same folder as \texttt{LHEReader.py}, the following terminal-based command will convert the input file to \Root format:

\begin{terminalbox}
python LHEReader.py --input input.lhe --output output.root
\end{terminalbox}

Alternatively, if the input and output files are in different paths, one can use: 

\begin{terminalbox}
python LHEReader.py --input <path_to_input>/input.lhe --output <path_to_output>/output.root
\end{terminalbox}

The input and output names of files can be replaced as per the user. The \texttt{output.root} is the name of the output \Root file produced. It is also possible to give the input as an \texttt{.lhe.gz} file, and the program will convert this into a \Root file. 

\begin{figure*}[htbp]
    \centering
    \includegraphics[width=0.98\linewidth]{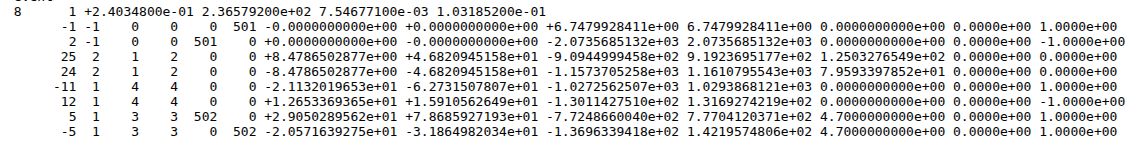}
    \caption{Screenshot of an event as stored in a sample \lhe file~\cite{Alwall:2006yp}. Each row corresponds to a single particle. The columns represent, from left to right: the PDG particle identifier, status code, first and second mother particle indices, first and second colour tags, four-momentum components ($p_x$, $p_y$, $p_z$, $E$) and mass, proper lifetime, and helicity.}
    \label{fig:lheevent}
\end{figure*}

An example of an event stored in the \lhe file is given in \autoref{fig:lheevent}. Each row corresponds to a single particle, storing its identity, kinematics, decay history, colour, and spin information.

We study the performance of the conversion on two sets of Monte Carlo simulations: $pp\to jj$ and $e^+e^- \to ZH$, both of which are simulated using \MG at the leading order. We simulated samples with a varying number of events to evaluate the average time it takes to convert a given event from \lhe format to \Root format. The results of this study are summarised in \autoref{tab:ppjj_times_ms} and \autoref{tab:eezh_times_ms}. We find that on average it takes about 0.20~ms (0.04~ms) to convert a given event from \lhe format to \Root format when $10^5$ events are generated for $pp\to jj$ ($e^+e^- \to ZH$).

\begin{table}[h!]
\centering
\begin{tabular}{ccc}
\toprule
\textbf{Events} & \textbf{Total Time (s)} & \textbf{Time/Event (ms)} \\
\midrule
$1\cdot 10^4$ & 2.7    & 0.27 \\
$5\cdot 10^4$ & 10.4   & 0.21 \\
$1\cdot 10^5$ & 19.7   & 0.20 \\
$1\cdot 10^6$ & 176.6  & 0.18 \\
\bottomrule
\end{tabular}
\caption{Average processing time for the Monte Carlo sample simulation of $pp\to jj$.}
\label{tab:ppjj_times_ms}
\end{table}

\begin{table}[h!]
\centering
\begin{tabular}{ccc}
\toprule
\textbf{Events} & \textbf{Total Time (s)} & \textbf{Time/Event (ms)} \\
\midrule
$1\cdot 10^4$ & 1.4   & 0.14 \\
$5\cdot 10^4$ & 2.4   & 0.05 \\
$1\cdot 10^5$ & 3.8   & 0.04 \\
$1\cdot 10^6$ & 28.4  & 0.03 \\
\bottomrule
\end{tabular}
\caption{Average processing time for the Monte Carlo sample simulation of $e^+ e^-\to ZH$.}
\label{tab:eezh_times_ms}
\end{table}

\section{\event function}\label{sec:event}

In this section, we illustrate an \event function written in \texttt{C++}~\cite{cpp} that enables one to process the \Root file obtained after following the commands illustrated in the previous section. We note that the parton-level analysis given here corresponds to using truth labels assigned by Monte Carlo event generators. The Particle Identification codes (PIDs) assigned to the leptons, neutrinos, and $b$-quark are summarised in \autoref{tab:pdgcode}. Anti-particles are assigned a negative PID. For other PIDs, we refer the reader to Ref.~\cite{ParticleDataGroup:2024cfk}. One can also access the status code of the particle, which indicates whether the particle is in the final state or is an intermediate particle.

\begin{table}[h!]
\centering
\caption{PDG codes for selected fundamental particles~\cite{ParticleDataGroup:2024cfk}.}
\begin{tabular}{lcc}
\toprule
\textbf{Particle} & \textbf{Symbol} & \textbf{PDG ID} \\
\midrule
Bottom quark & $b$ & 5 \\
Electron & $e^-$ & 11 \\
Muon & $\mu^-$ & 13 \\
Neutrinos & $\nu_e, \nu_\mu, \nu_\tau$ & 12, 14, 16 \\
Higgs boson & $H$ & 25 \\
Z boson & $Z^0$ & 23 \\
\bottomrule
\end{tabular}
\label{tab:pdgcode}
\end{table}

The content of the \lhe file is stored as a \texttt{TTree} object named \texttt{events} in the \Root file. The various branches stored in this \Root file and their types are summarised in \autoref{tab:branch}.

\begin{table}[htbp]
\centering
\caption{Summary of branches stored in the \texttt{events} \texttt{TTree}.}
\begin{tabular}{lcl}
\toprule
\textbf{Category} & \textbf{Branch} & \textbf{Type} \\
\midrule
Event & \texttt{numParticles} & \texttt{int} \\
Event & \texttt{eventweight}  & \texttt{float} \\
Event & \texttt{scale}        & \texttt{float} \\
Event & \texttt{alpha\_qed}   & \texttt{float} \\
Event & \texttt{alpha\_qcd}   & \texttt{float} \\
\midrule
Particle & \texttt{pid}     & \texttt{vector<int>} \\
Particle & \texttt{status}  & \texttt{vector<int>} \\
Particle & \texttt{mother1} & \texttt{vector<int>} \\
Particle & \texttt{mother2} & \texttt{vector<int>} \\
Particle & \texttt{color1}  & \texttt{vector<int>} \\
Particle & \texttt{color2}  & \texttt{vector<int>} \\
Particle & \texttt{px}      & \texttt{vector<float>} \\
Particle & \texttt{py}      & \texttt{vector<float>} \\
Particle & \texttt{pz}      & \texttt{vector<float>} \\
Particle & \texttt{energy}  & \texttt{vector<float>} \\
Particle & \texttt{mass}    & \texttt{vector<float>} \\
Particle & \texttt{tau}     & \texttt{vector<float>} \\
Particle & \texttt{spin}    & \texttt{vector<float>} \\
\bottomrule
\end{tabular}
\label{tab:branch}
\end{table}

A simple method to access these branches in \Root is as follows:

\begin{codebox}
TFile *f = new TFile("in.root","read");
TTree *t = (TTree*)f->Get("events");
Float_t         eventweight;
vector<int> *pid = 0;
t->SetBranchAddress("eventweight", &eventweight);
t->SetBranchAddress("pid", &pid);
\end{codebox}

The event loop itself can be initiated in the following way:

\begin{codebox}
for(Int_t entry=0; entry < total_entries; ++entry){ // event loop
t->GetEntry(entry);
for(Int_t i=0; i < pid->size(); ++i){ // particle loop
}// end particle loop
} // end event loop
\end{codebox}

To access final-state particles (status code~$= 1$), one can use the following within the particle loop:

\begin{codebox}
if(status->at(i) == 1){
<script>;
} 
\end{codebox}

Access to an individual particle is provided through its particle identification (PDG code):

\begin{codebox}
if(abs(pid->at(i))==5){
<run script on b-jet>;
}
\end{codebox}

To compute the missing transverse energy, the following is used:

\begin{codebox}
if(abs(pid->at(i))==12 || abs(pid->at(i))==14 || abs(pid->at(i))==16){
MET.SetPxPyPzE(px->at(i),py->at(i),pz->at(i),e->at(i));
myMET.SetPtEtaPhi(MET.Pt(),0.0,MET.Phi());
myMET.SetZ(0.0);
}   
\end{codebox}

Selections can be applied using the following conditional statement:

\begin{codebox}
if(myMET.Pt() < 25){continue;}
\end{codebox}

where the events with $E_T^{\rm miss} < 25$~GeV are rejected.

An \texttt{EventLoop.cpp} example implementing the patterns above is provided in the \lhereader repository~\cite{aman_desai_2026_18822147}. It can be executed directly using \Root as:
\begin{terminalbox}
root -l -q -b EventLoop.cpp
\end{terminalbox}
The code reads input \Root files from the \texttt{samples/} directory and writes normalised histograms to \Root output files and PDF plots to the \texttt{plots/} directory. A \texttt{README.md} in the repository provides further usage instructions.

% ---------- Switch to one column ----------

\begin{figure*}[htbp]
    \centering
    \vspace{.3cm}
    \includegraphics[width=0.45\linewidth]{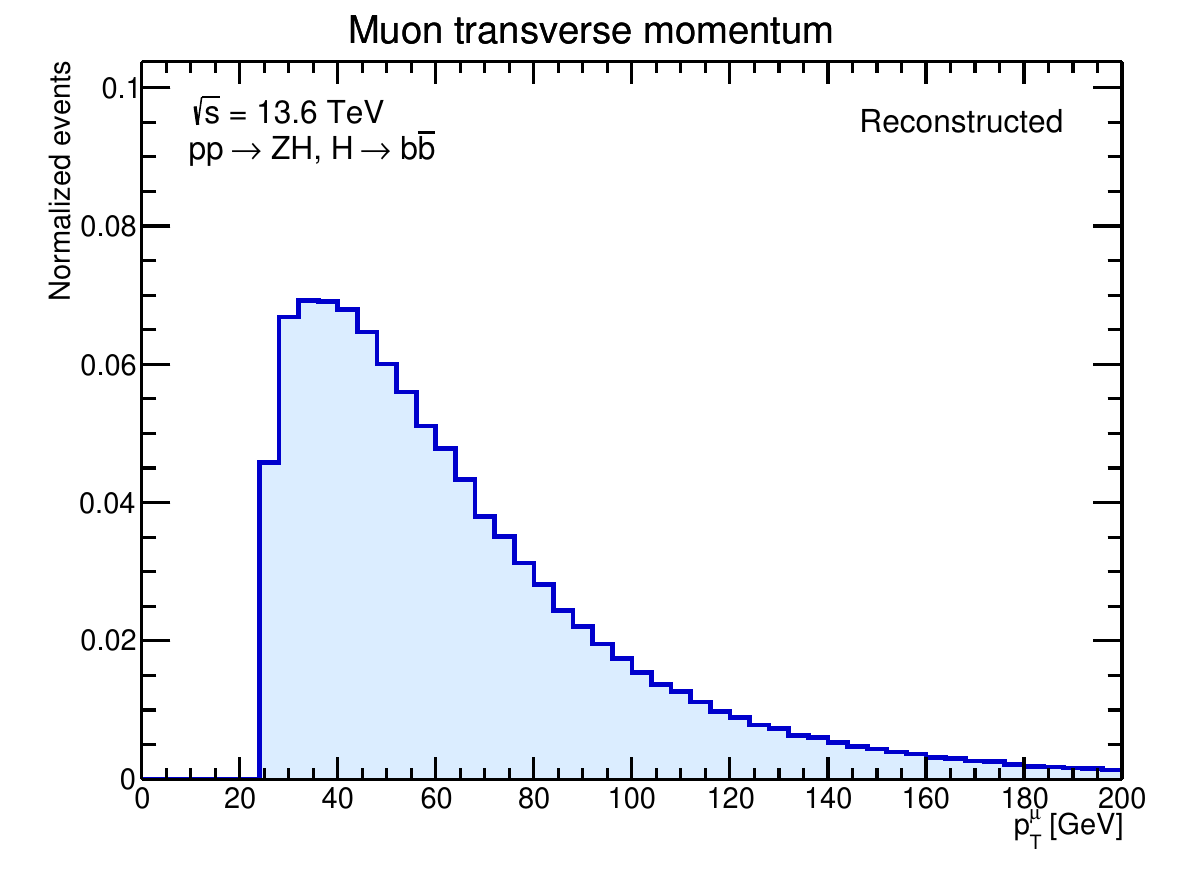}
    \includegraphics[width=0.45\linewidth]{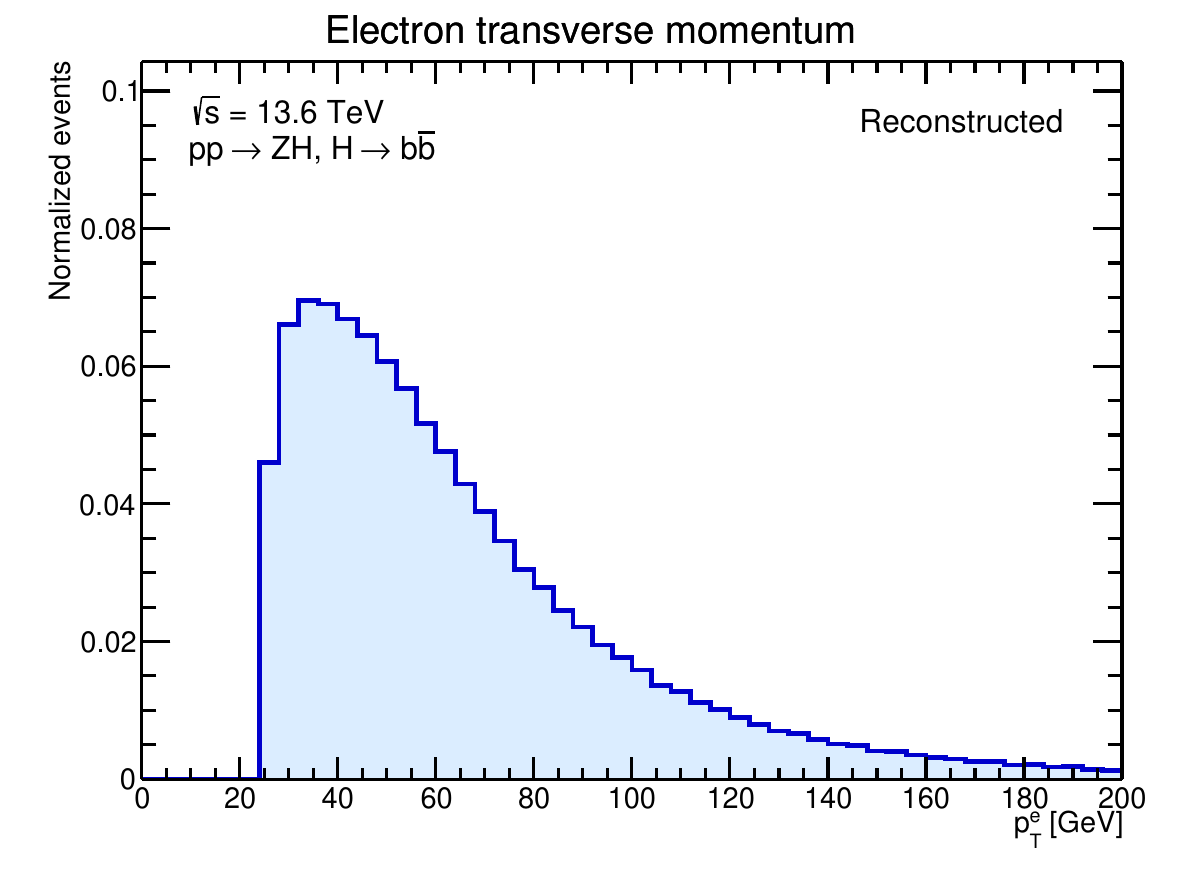}\\
    \includegraphics[width=0.45\linewidth]{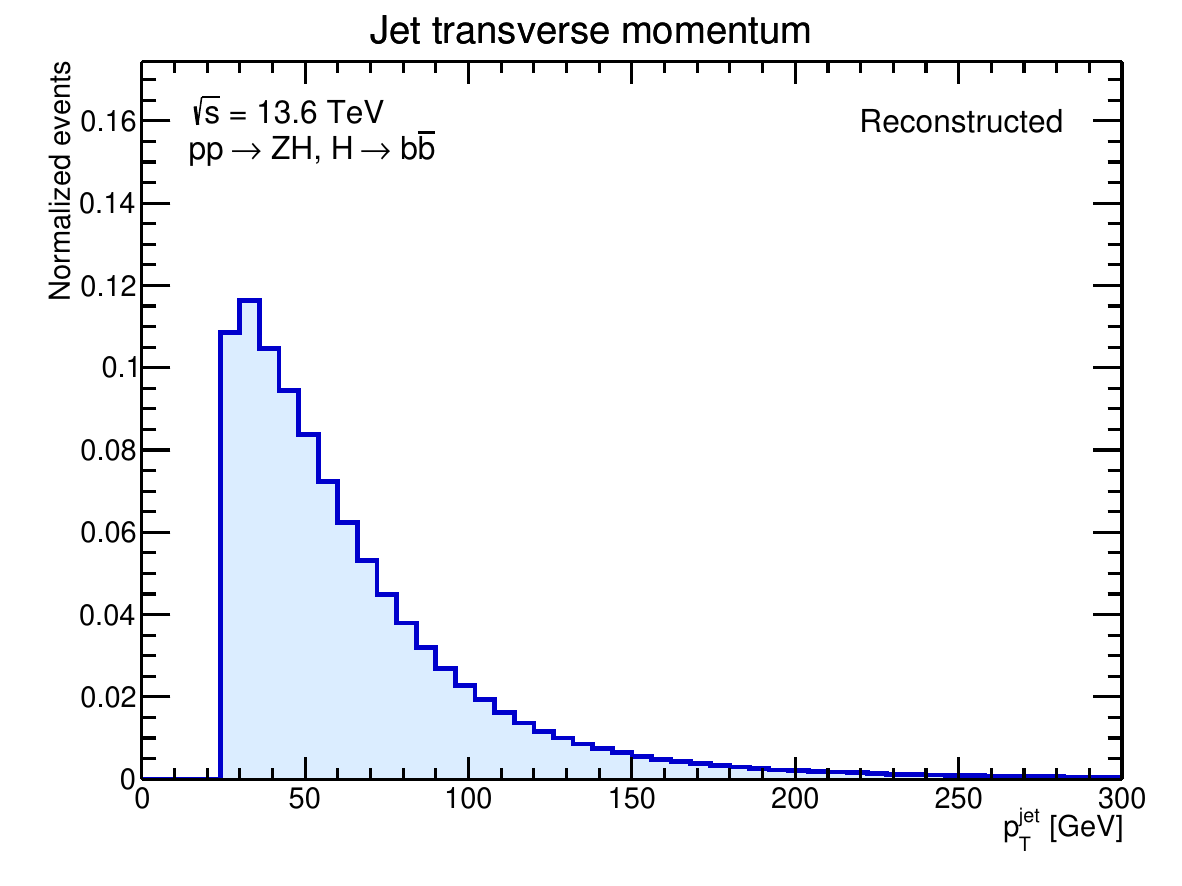}
    \includegraphics[width=0.45\linewidth]{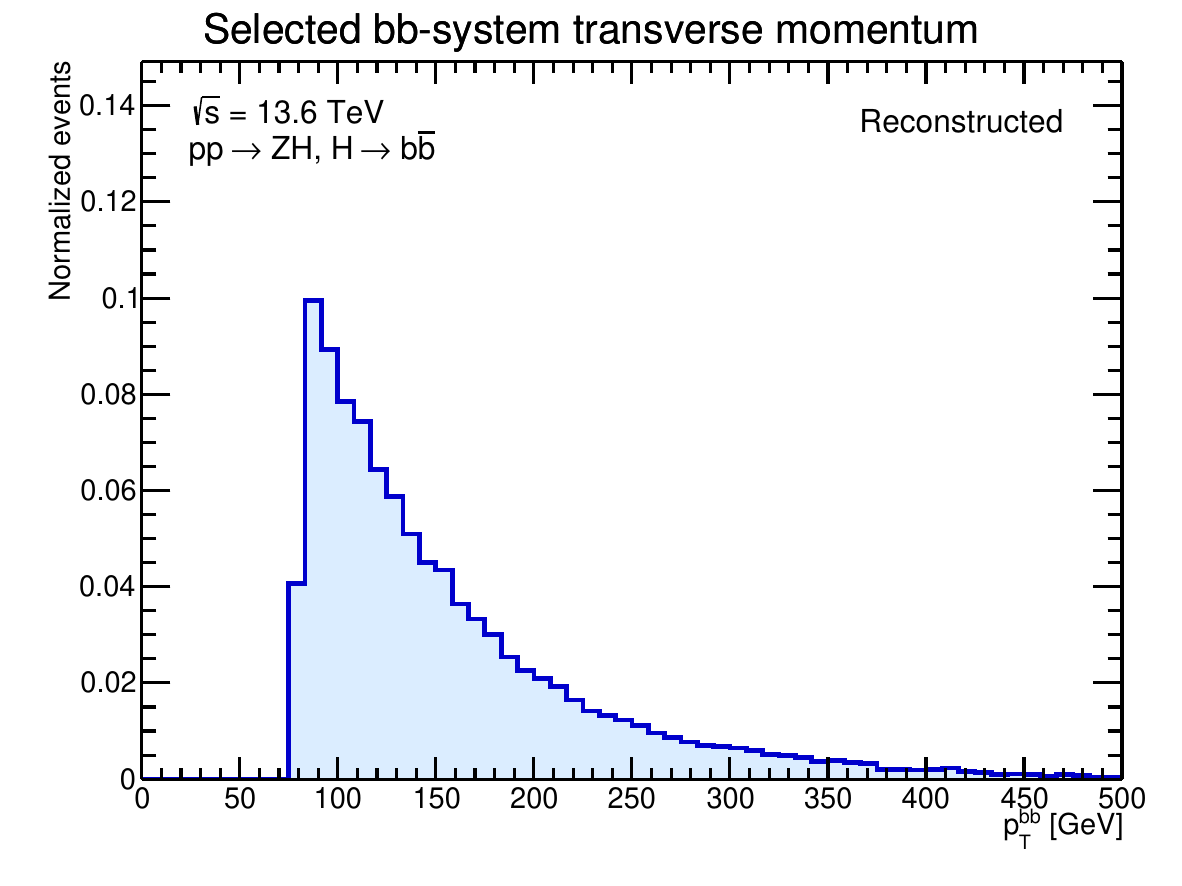}
    \caption{
    Distributions of selected kinematic observables at the reconstructed level:
    muon transverse momentum,
    electron transverse momentum,
    $b$-jet transverse momentum,
    and the transverse momenta of the pair of $b$-jets.
    }
    \label{fig:kinematic1}
\end{figure*}

\begin{figure*}[htbp]
    \centering
    \includegraphics[width=0.45\linewidth]{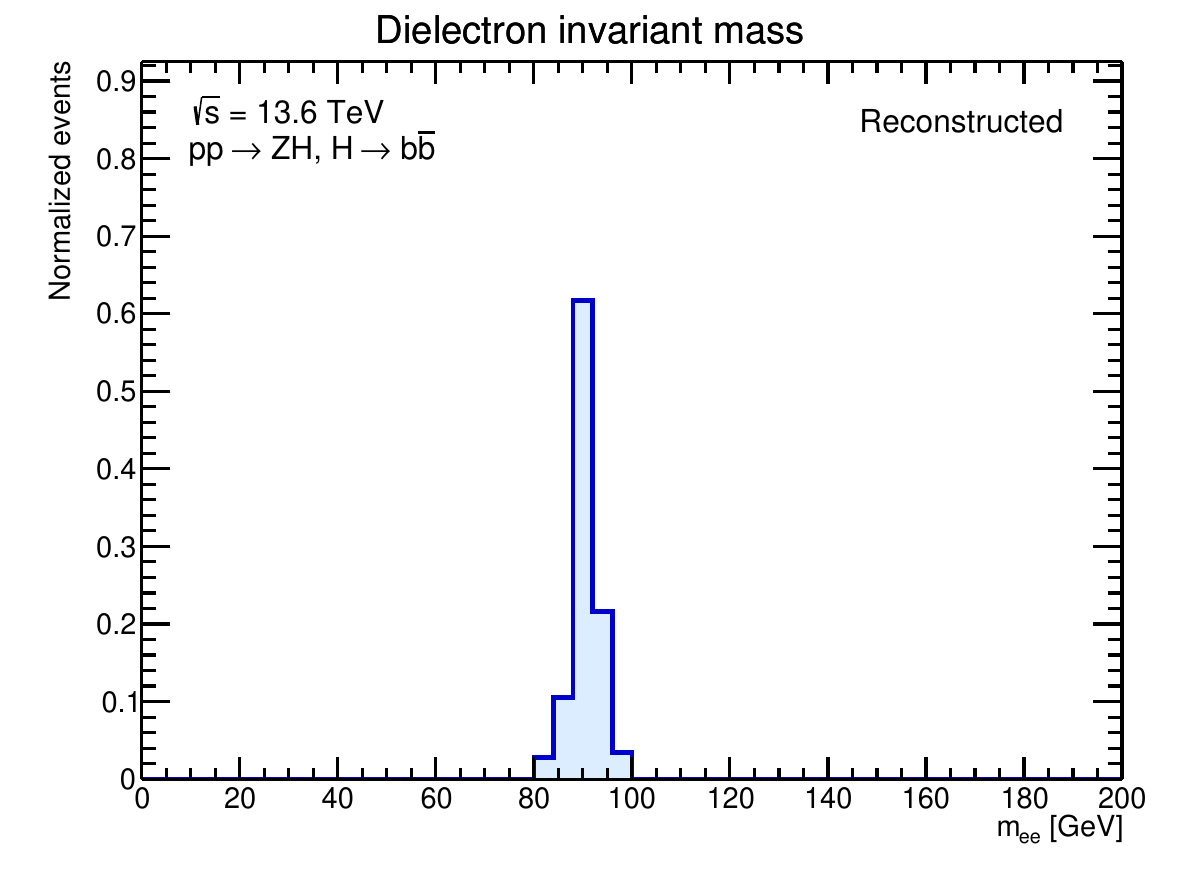}
    \includegraphics[width=0.45\linewidth]{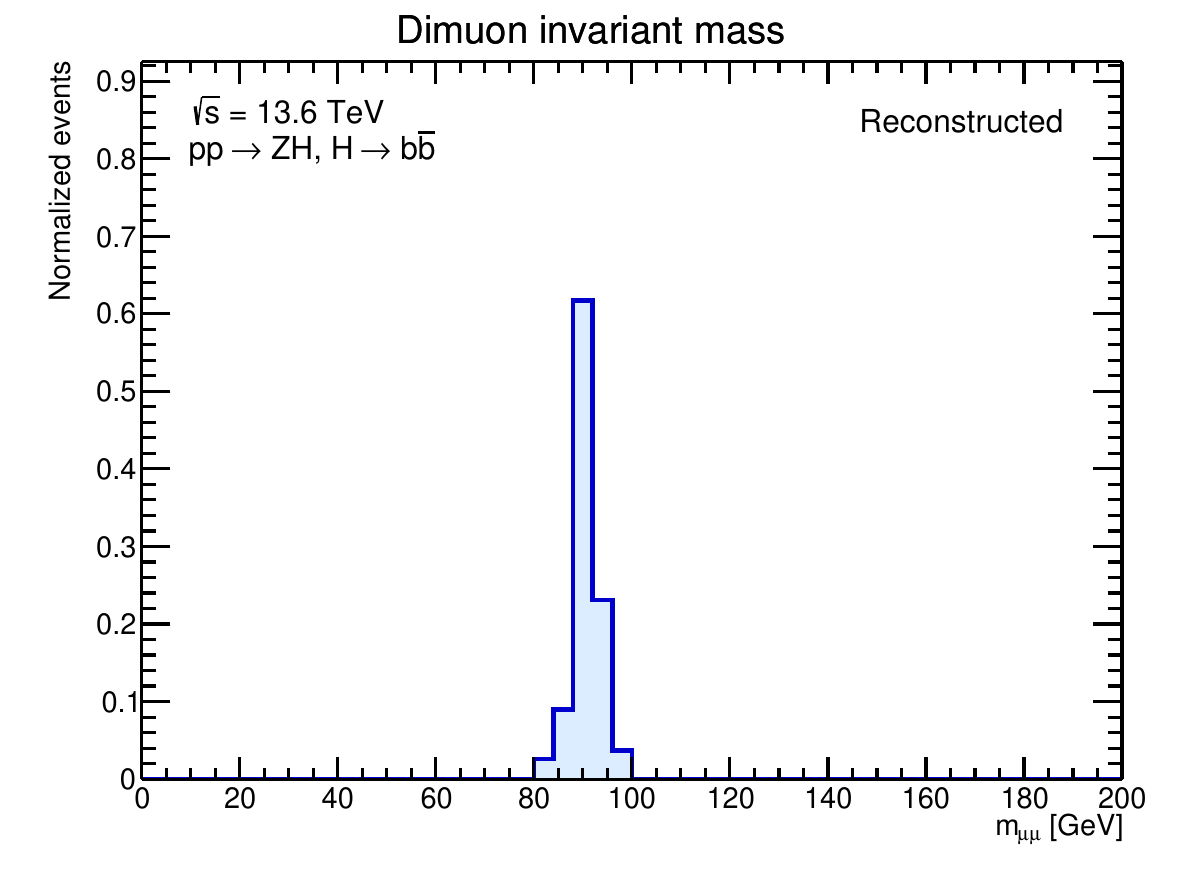}
    \caption{
    Di-lepton invariant mass distributions:
    $m_{ee}$ (left) in the two-electron channel and
    $m_{\mu\mu}$ (right) in the two-muon channel.
    }
    \label{fig:kinematic2}
\end{figure*}

\begin{figure*}[htbp]
    \centering
    \includegraphics[width=0.45\linewidth]{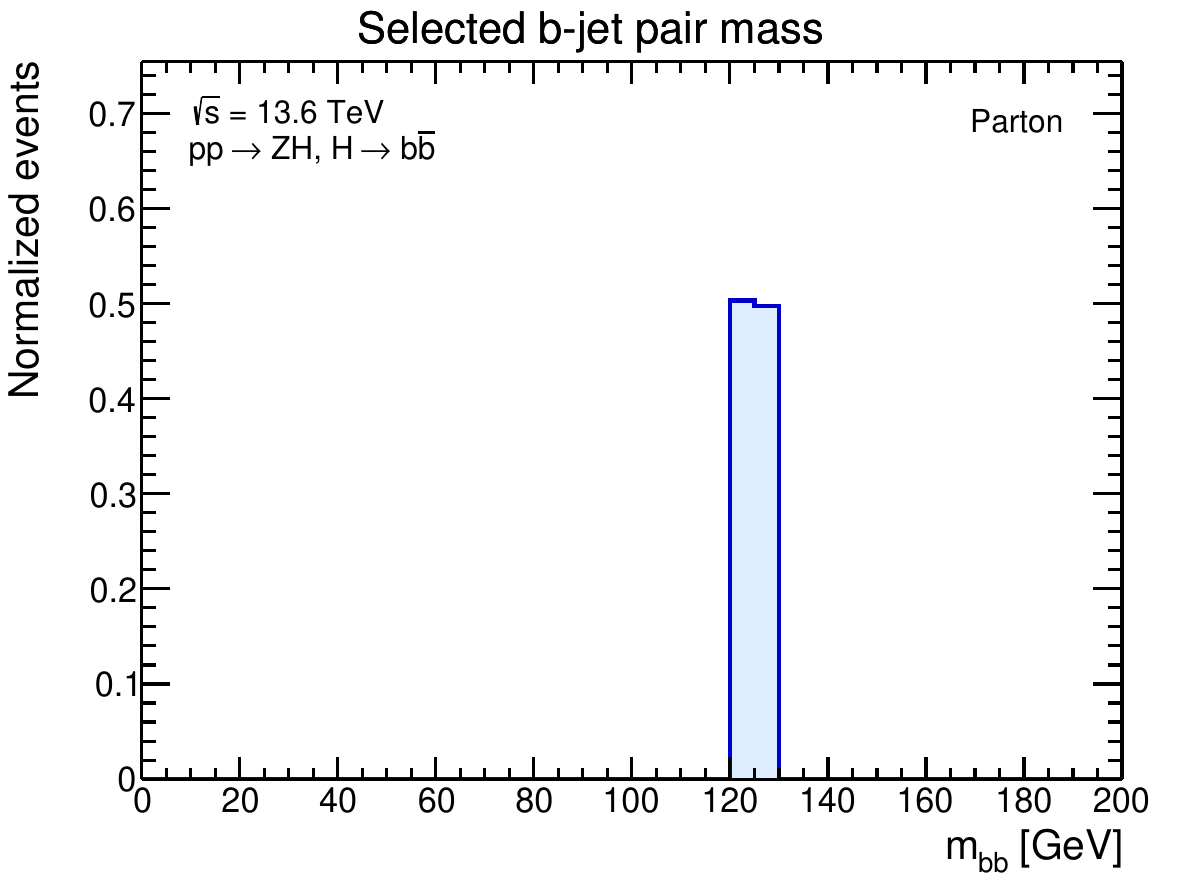}
    \includegraphics[width=0.45\linewidth]{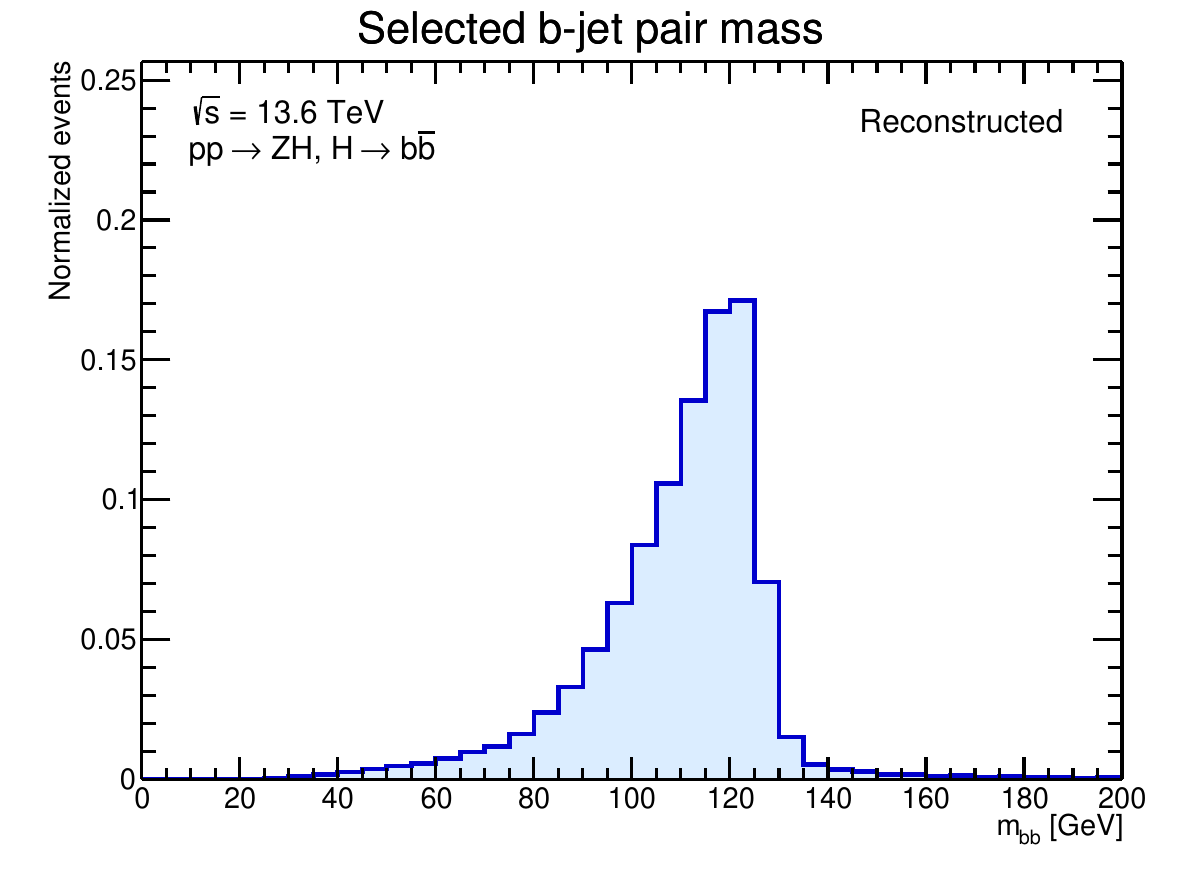}
    \caption{
    Higgs boson mass distribution at parton level (left) and reconstructed level (right).
    }
    \label{fig:higgsmass}
\end{figure*}

% ---------- Back to two columns ----------
\section{Application}\label{sec:mcsample}

We illustrate the application of this program using an example of Monte Carlo events produced using \MG~\cite{Alwall:2014hca}. We simulate the $pp \rightarrow Z H$ process with the $Z$ boson decaying to a pair of leptons (electrons or muons) and the Higgs boson decaying to a pair of $b$-quarks. With this example, we not only illustrate the usage of \lhereader but also the \event function to process quarks, leptons, and missing energy. A representative Feynman diagram corresponding to the process is given in \autoref{fig:ZH_diagram}. We set the beam energy to 6.8 TeV corresponding to LHC Run 3 and use the PDF4LHC21 PDF set~\cite{PDF4LHCWorkingGroup:2022cjn}, which is accessed using \textsc{Lhapdf6}~\cite{Buckley:2014ana}.  The cross-section including the decay branching fractions for this process, as obtained from \MG, is $39.6$~fb. The parton-level events are then processed through \Pythia~\cite{Bierlich:2022pfr} for parton shower and hadronisation. Subsequently, we process these through \FastJet~\cite{Cacciari:2011ma} for clustering objects into jets, through its interface as provided within \MadAnalysis~\cite{Conte:2012fm,Conte:2018vmg}. We have used the anti-$k_T$ algorithm~\cite{Cacciari:2008gp} with the radius parameter set to 0.4. The results are presented after normalisation.

A simplified analysis strategy is adopted to demonstrate the application of \lhereader and the \event function. We select events consisting of exactly two opposite-sign, same-flavour isolated leptons, each satisfying the requirements $p_T>25$~GeV and $|\eta| < 2.5$. Moreover, the invariant mass of the leptons is required to be within $[80,100]$~GeV, to select events consistent with $Z\to \ell^+\ell^-$. We require jets with $p_T>25$~GeV and $|\eta| < 2.5$. We select $b$-jets using truth information for parton-level analyses and use the PID branch for reconstructed-level analysis. The selected $b$-jets and leptons are required to have $\Delta R(b,\ell) > 0.4$. At least two $b$-jets are required to be present in an event. The leading and sub-leading $b$-jets, with a combined transverse momentum greater than 30~GeV, are used to reconstruct the Higgs candidate.

Selected distributions of kinematic features after the reconstruction stage are summarised in \autoref{fig:kinematic1}. The distributions correspond to the transverse momenta of the muons, electrons, and $b$-jets. In \autoref{fig:kinematic2}, we show the distributions of the $Z$ boson mass candidate reconstructed from a pair of selected electrons and muons. The Higgs boson mass distribution, at the parton level, is a single bin at the Higgs boson mass, and at the reconstructed level is spread about the Higgs boson mass as shown in \autoref{fig:higgsmass}.

\section{Conclusions}

We presented \lhereader, which reads and converts a given \lhe file into a \Root file. The \lhereader program requires only \Root (with \textsc{PyRoot} enabled), while the other dependencies are available in the \textsc{Python} installation. The program accepts both \texttt{.lhe} and \texttt{.lhe.gz} as input files, producing a standard \texttt{TTree}-based \Root output. We evaluated the conversion performance on simulated $pp\to jj$ and $e^+e^-\to ZH$ samples, finding an average conversion time of approximately 0.20~ms and 0.04~ms per event, respectively, for $10^5$ generated events. We also presented an example \texttt{C++} \event code that can be used to read the \Root file. As an example, we showed how this program is applied to the $pp\to ZH$ process, with $Z\to\ell^+\ell^-$ and $H\to b\bar{b}$, using the \lhe file output generated by \MG and by \MadAnalysis. The scripts presented in this paper could be useful for research purposes, providing a quick way to test kinematic distributions and the \lhe file format, as well as for educational purposes, serving as a lightweight tool to convert an \lhe file to a \Root file. In the future, we aim to extend \lhereader to support use as an importable Python module in addition to its existing standalone command-line execution.

\bibliographystyle{JHEP}
\bibliography{biblio}

\end{document}